\documentstyle[12pt,epsfig]{article}

\newcommand{\gev}{ {\rm GeV} } 
\newcommand{\tev}{ {\rm TeV} } 
\newcommand{\mm}{ {\rm mm} } 
\newcommand{\cm}{ {\rm cm} } 
\newcommand{\mim}{ {\mu \rm m} } 
\newcommand{\be}{\begin{equation}}
\newcommand{\ee}{\end{equation}}
\newcommand{\bea}{\begin{eqnarray}}
\newcommand{\eea}{\end{eqnarray}}
\newcommand{\ba}{\begin{array}}
\newcommand{\ea}{\end{array}}

\newcommand{\nn}{\nonumber}
\begin{document}
%
\renewcommand{\thefootnote}{\alph{footnote}}
\begin{titlepage}
\vspace*{-1cm}
\phantom{bla}
\vskip 2.0cm
\begin{center}
{\Large \bf Winding Modes and Large Extra-Dimensions}
\end{center}
\vskip 1.5cm
\begin{center}
{\large A. Donini and S. Rigolin} \\
\vskip .1cm
{{\it Departamento de F\'{\i}sica Te\'orica, Universidad Aut\'onoma de Madrid, 
SPAIN} \\
 {\rm E--Mail: donini@daniel.ft.uam.es, rigolin@delta.ft.uam.es} }
\end{center}
\vskip 3.0truecm
%
\begin{center}
\bf Abstract
\end{center}
%
\begin{quote}
We review briefly the main features of the Large Extra Dimensions
scenario in the framework of weakly coupled Type I string theory.
Kaluza-Klein (KK) excitations of the graviton are expected, whereas no
KK modes for the gauge bosons arise if the gauge group is tied to a
D3-brane. In this scenario, typical signatures such as direct
production of KK modes of the graviton at high-energy colliders could
test the size of the compactified dimensions. We point out that 
contrary to what considered in the literature on the subject, in the general
case of anisotropic compactification Winding Modes of the Standard
Model gauge bosons could also be directly observable, thus further
constraining the model. 
\end{quote}
%
%
\vfill{Talk given at the XXXIVth Rencontres de Moriond by A. Donini.}
\end{titlepage}
\setcounter{footnote}{0}
\renewcommand{\thefootnote}{\arabic{footnote}}
\baselineskip=18pt
%
\subsection*{Large Extra Dimensions}
%
The Newton law is tested at present from 
very large scale down to $R_{exp} \ge 1 \ \cm$. However, 
the typical length scale where gravity is expected to become strongly
interacting, thus requiring the inclusion of quantum effects, is
$R_{Planck} \simeq 10^{-33} \cm$. It is possible that deviations from
the Einstein-Newton theory could be observed at the planned
gravitational experiments \cite{newexp} testing 4-dimensional
gravity down to $R \sim 10 \ \mim $.

In \cite{add} a scenario was proposed where extra dimensions 
open at the sub-millimeter scale, thus modifying the 
4-dimensional gravity. The relation between the fundamental 
$(4+n)$-dimensional Planck scale and the 4-dimensional one is
\be
\label{eq1}
\frac{M_{Planck}^2}{8 \pi} = R^n M_{(4+n)}^{2+n}
\ee
where $R$ is the size of the compactified dimensions.
If the compactified volume ($V \sim R^n$) is large enough the
fundamental Planck scale $M_{(4+n)}$ could be well below the
4-dimensional one, $M_{Planck} \sim 10^{19} \ \gev$.
If the fundamental Planck mass is at the TeV-scale, 
the long-standing hierarchy problem is solved in a natural way: 
in this framework, there is no
hierarchy at all (or a very small one) between the electroweak scale 
and the fundamental Planck scale. 
New physics at the TeV-scale would be quantum gravity itself. 

Since the only consistent theory of quantum gravity
at present implies an underlying (10-dimensional) superstring theory 
and the string scale is related to the fundamental Planck scale, 
the scenario proposed in \cite{add} naturally involves
string theory at the TeV-scale, such as was soon noticed in 
\cite{aadd}. The string scale is directly fixed to the
4-dimensional Planck scale in weakly coupled
heterotic string theory, $M_s = \frac{\sqrt{\alpha}}{2} M_{Planck}$
(where $\alpha$ is the GUT gauge coupling), and thus it is
not possible to lower it to the TeV-scale. In weakly coupled 
Type I string theory, however, the string scale is not
uniquely fixed by the 4-dimensional Planck scale 
and the gauge coupling, the relation involving also the 
compactified volume $V$. It is then possible to achieve a low-energy 
string scale (see \cite{string} for details on heterotic and Type I strings). 
The scenario proposed in \cite{add,aadd}, then, implies
the exciting possibility of direct experimental observation
of stringy effects at the planned high energy colliders, such as LHC.
%
\subsection*{D-branes and String Modes}
%
Due to the compactification of extra dimensions,
gravitons and Standard Model particles acquire
massive replicas, the Kaluza-Klein (KK) excitations, with mass
$m_{KK} \propto 1/R$. If the radius is large
enough to bring down the string scale to 1 TeV, these massive excitations
are extremely light. If the compactification radius is $\sim 1 \ \mm$, 
$m_{KK} \sim 10^{-12} \ \gev$.
Clearly, such a low mass for SM particles replicas is forbidden
by the experiments. A possible solution to this problem can be found
by considering the SM particles confined on a D3-brane, whereas
gravity lives in the full 10-dimensional space-time \cite{aadd}.

D$p$-branes are extended objects with $p$ spatial dimensions
and appear naturally in many string theories
as classical solutions of 10-dimensional gravity.
In the classical limit, they can be naively viewed 
as static surfaces with open strings tied to, see \cite{polrev}.
Compactifying on a $T^2 \times T^2 \times T^2$ manifold 
the 10-dimensional effective action for weakly coupled Type I strings,
we get
\bea
S_{4} \ &  = & \ - \int {{ d^4 x}\over {2\pi }}
\sqrt{-g}\ \left(\ {{R_1^2 R_2^2 R_3^2 M_s^8}\over{\lambda^2}} \ {\cal
R} \ + \  
{{R_1^2 R_2^2 R_3^2 M_s^6}\over{\lambda}} \ {1\over 4} F^2_{(9)}\right. \nn \\ 
& + & \left. \ \sum _{i\not= j\not=k\not= i} {{R_i^2 R_j^2 M_s^4}\over 
{\lambda}} \ {1\over 4} F^2_{(7_k)} \ + \ \sum_{j=1}^3 
{{R_j^2 M_s^2}\over {\lambda}} \ {1\over 4}  F^2_{(5_j)} \ +
{1\over {\lambda}}\ {1\over 4}  F_{(3)}^2  \ +\ ...\  \right)\ ,
\label{d04}
\eea
where $\lambda$ is the string coupling, $M_s$ the string scale, 
${\cal R}$ is the trace of the Ricci tensor (coming from the closed
string sector) and $F_{(3)},F_{(5)},F_{(7)}$ and $F_{(9)}$ represent the gauge 
field strengths associated to the 3-, 5-, 7- and 9-brane sector
respectively. These gauge fields can be imagined as the ending points
of open strings tied to extended objects with 3, 5, 7 and 9 
spatial dimensions. The compactified volume is $V = \Pi_i (2 \pi R_i)^2$
and $R_i$ are the radii of the three complex tori \cite{imr}.

The relation between the string scale, the 4-dimensional Planck scale
and the gauge coupling is (for the 3-brane gauge group)
\be 
\frac{\alpha_3 M_{Planck}}{\sqrt{2}} 
    = \left( M_s^4 R_1 R_2 R_3 \right)
\label{gaugnos}
\ee
and changing the size of the compactified manifold the string scale
can be much lower than the Planck scale.
If we assume that gauge SM particles are massless excitations (zero 
modes) of open strings starting and ending on, say, a 3-brane,
at a first look it seems that gauge fields cannot see the extra dimensions, 
since these ending points cannot move freely in the dimensions
transverse to the brane. This is not actually true, as will become 
clear by looking at the spectrum of string modes.

Consider first a closed string, whose zero-modes represent the graviton.
The mass squared of a given excitation is:
\be
\label{closedspec}
M_{closed}^2(m_i, n_i) = M_0^2 + \sum_{i=1}^3 
\left( m_i^2 \frac{1}{R^2_i} + n_i^2 (M_s^2 R_i)^2 \right) \ ,
\qquad m_i, \  n_i = 0, \pm 1, \pm 2, \dots
\ee  
where $M_0$ stands for $R_i$-independent contributions to the mass
proportional to the string scale and $(m_i, n_i)$ are the Kaluza-Klein
and winding numbers. These two kinds of excitations are related to the
compactification from 10- to 4-dimensional space-time: 
KK modes represent the quantization of momenta in the compactified
dimensions of the string; winding modes represent the wrapping around 
the compact dimension of the string (and $M_{\omega_i} = M_s^2 R_i$). 
All the possible combinations of these
two quantum numbers are possible for a closed string, since 
a closed string wrapped around the torus 
is topologically different from an unwrapped one.

For an open string the mass squared of a given excitation is:
\be
\label{openspec}
M_{open}^2(m_i) = M_0^2 + \sum_{i=1}^3 \left( m_i^2 \frac{1}{R^2_i} \right) .
\ee  
Only KK modes are possible for an open string, since an open string
wrapped around the compact dimension is topologically equivalent
to an unwrapped string.

Finally, for a D3-brane, we have
\be
\label{3branespec}
M_{3-brane}^2(n_i) = M_0^2 + \sum_{i=1}^3 
\left( n_i^2 (M_s^2 R_i)^2 \right)
\ee  
and no KK modes are possible, since the endings of the open string are 
tied to the non-compact (3+1)-dimensional space-time and they cannot
move into the compact dimensions. Notice, however, that 
such an open string wrapped around the compact dimension is
topologically different from an unwrapped one, thus reintroducing
winding modes in the spectrum.

If we consider a very large compact radius 
the KK modes are light whereas the winding modes are heavy
and decouple from the low-energy spectrum. Moreover, if the 
string scale is still high enough, also terms proportional to $M_0^2$  
decouple. In this framework, we can consider that SM particles have no
KK replicas and only the graviton does. Winding modes of SM particles 
and of the graviton are heavy and are not relevant for low-energy
physics.
%
\subsection*{Winding Modes Phenomenology}
%
It has been mentioned that planned gravitational experiments
could test the Newton law down to $\sim 10 \ \mim$, thus
putting an upper bound on the size of the compactified manifold
and therefore on the string scale. It is also possible that 
new high-energy experiments at the $\tev$-scale such as LHC or NLC 
could directly observe the production of KK modes of the graviton \cite{grw}.
Although graviton emission in 4-dimensional quantum gravity 
is suppressed by $1 / M_{Planck}^2$, the integration 
over all the KK modes of the graviton trades this suppression 
factor with a much smaller one, 
\be
\frac{1}{M_{Planck}^2} \to \frac{E^n}{M_{(4+n)}^{2 + n} } ,
\ee
where $n$ is the number of dimensions with large radius, $M_{(4+n)}$ is the
fundamental Planck scale and $E$ is the c.m. energy. 
If the $(4+n)$-dimensional Planck mass is at the $\tev$-scale, 
graviton production becomes sizeable at the planned accelerators. 
The typical signature will be the production of a SM particle 
and missing energy \cite{grw}. 

If the compactification radius is $R \sim 1 \ \mm$ and $M_{(4+n)} \sim 
1 \ \tev$, this scenario can be tested at the planned gravitational
and high-energy experiments. Slightly changing these scales 
dramatically changes its experimental appeal for the near future. 
In particular, the number of extra dimensions felt by gravity
at the $\tev$-scale and the precise values of $M_{(4+n)}$ and of the 
compactification radii, $R_i$, are fundamental when making quantitative
predictions on the decay rate into gravitons or on a specific cross-section. 

In the isotropic case $n=6 \ (R_1 = R_2 = R_3)$ we face two possibilities:
\begin{itemize}
\item
We observe deviations from the Newton law at the planned gravitational
experiments, $R \sim 1 \ \mm$. In this case, we immediately get 
$M_s \sim 10^{-5} \ \gev$ (for $\alpha_3 = 1/24$). 
Clearly, this situation is excluded by experiments, 
since the string scale is too low.
\item
$M_s \sim 1 \ \tev$. In this case we get $M_c \sim 10^{-2} \ \gev$
(i.e. $R \sim 10^{-11}$ mm), and thus no extra dimensions could be
observed at the planned gravity experiments. 
\end{itemize}
In the literature, it has been shown that the case with $n=2$ 
gives the favoured signatures at the planned experiments.  
For $n=2$ large dimensions ($ R_1 > R_2 \sim R_3 $),
it is possible to have simultaneously $M_s \sim 1 \ \tev$ and 
$R_1 \sim 0.1 \ \mm$, thus observing deviations from the Newton law
and graviton emission into the bulk at the same time.

This simple analysis shows that, from the experimental point of view, 
the most interesting scenario is represented by an anisotropically
compactified manifold, with some large dimensions and some small ones.
Although the isotropic $n=6$ case is not excluded, 
its lower phenomenological interest with respect to the $n=2$ case is 
manifest. 

If we consider a large anisotropy, $R_1 >> R_2 \sim R_3$, 
the spectrum of the model
changes with respect to the case of three equally large radii considered
above. From eq. (\ref{3branespec}) we can see that the winding
modes of SM particles along the large compact dimension actually
decouple, whereas the winding modes along the small compact dimensions
do not. This was first noticed in \cite{dr}, pointing out that
in this case the light winding modes can play a phenomenological role.
Using eq. (\ref{gaugnos}) with $R_1 = 1 \ \mm$
and a gauge coupling $\alpha_3 = 0.1$, we get for the light winding modes
mass $M_{\omega} \sim 1 \ \tev$ (the precise value depending
on the gauge coupling $\alpha_3$ and hence on all the details 
of the specific model). Combining eq. (\ref{eq1}) and
eq. (\ref{gaugnos}) we get ($R_2 = R_3 = R_c$):
\bea
M^2_{(6)} = & \frac{1}{\sqrt{4 \pi} \alpha_3 } M^2_{\omega_c} 
& (R_1 > R_c) \nn \\
M^3_{(8)} = & \frac{1}{\sqrt{4 \pi} \alpha_3 } M^2_s M_{\omega_1} 
& (R_1 < R_c) \\
M^4_{(10)} = & \frac{1}{\sqrt{4 \pi} \alpha_3 } M^4_s 
& (R_1 = R_c) \nn 
\label{neweq}
\eea
respectively for $n=2, 4, 6$ Large Extra Dimensions.
\begin{figure}[t]
\begin{center}
\epsfig{file=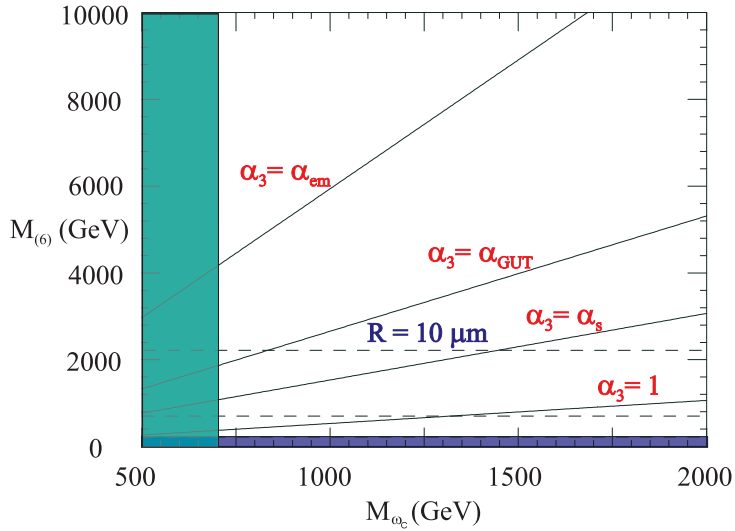,height=7cm} 
\end{center}
\caption{\it \small Bounds on $M_{(6)}$ as a function of 
$M_{\omega_c}$ for different values of $\alpha_3 = \alpha_{em}, \alpha_{GUT}, 
\alpha_s, 1$. 
The horizontal lines are the excluded region from present 
and planned gravitational experiments testing the Newton law down to
$R \sim 1 \ \cm, 1 \ \mm, 10 \ \mim$. The vertical 
shaded area is the excluded region from non observation of 
massive replicas of SM bosons.
}
\label{boundm6}
\end{figure}
Therefore, in the case $n=2$ for which the most promising
experimental signatures are foreseen, the relevant scale in the game
is the light winding mode mass $M_{\omega_c}$. This scale is directly related
to the 6-dimensional Planck mass, for which
bounds can be extracted from the experiments (see \cite{grw}). Notice that 
in this particular case the string scale is completely irrelevant from 
the phenomenological point of view and could take any value.

\begin{figure}[t]
\begin{center}
\epsfig{file=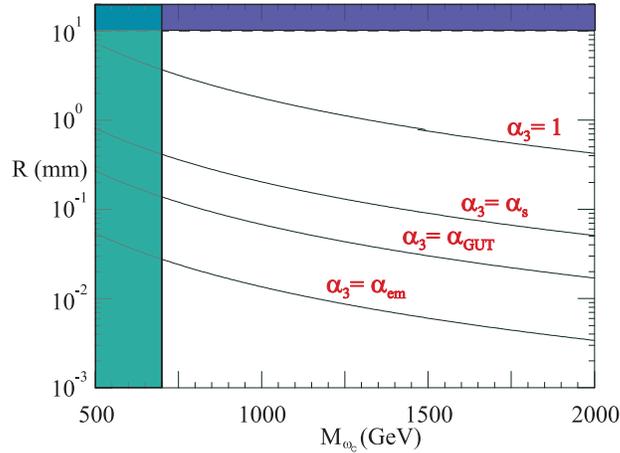,height=6cm} 
\end{center}
\caption{\it \small Bounds on $R_1$ as a function of $M_{\omega_c}$ for 
different values of $\alpha_3 = \alpha_{em}, \alpha_{GUT}, \alpha_s, 1$.
The upper shaded area is the excluded region from present gravitational 
experiments. The vertical shaded area is the excluded region 
from non observation of massive replicas of SM bosons.
}
\label{boundr}
\end{figure}

Existing experimental limits on massive replicas of SM gauge bosons, 
such as $Z'_{SM}$ or $W'_{SM}$, can be directly translated into 
$M_{\omega_c} \ge 700$ GeV \cite{D0CDF}.
This bound can be used to put limits on $R_1, M_{(6)}$ and $M_s$.
For $n=2$, with $\alpha_3 = \alpha_{GUT} = 
1/24$, we obtain the limit $M_{(6)} \ge 1.8$ TeV.
Contextually, using eq.~(\ref{eq1}) we get $R \le 0.15 \ \mm$.
Non-observation of massive replicas of SM particles at the planned accelerators
would imply even stronger bounds on $M_{(6)}$. For example, non-observation 
at NLC500 (i.e. $M_{Z'_{SM}} \ge 5$ Tev), translates into 
$M_{(6)} \ge 13$ TeV and $R \le 3 \times 10^{-3} \ \mm$.
The dependence of the previous limit on $\alpha_3$ 
is shown in Fig.~\ref{boundm6} and Fig.~\ref{boundr}. 

Search for direct production of winding modes at LHC or NLC
represents a useful tool to explore the parameter space of 
Large Extra Dimension models derived from Type I string theory.

%
We thank A. de Rujula, M. B. Gavela, L. Ib\'a\~nez and C. Mu\~noz
for useful discussions. 
%

%
\end{document}